\documentclass[prc,aps,showpacs,showkeys,groupedaddress,floatfix,superscriptaddress]{revtex4-1}

\usepackage{rotating}

\usepackage{epsfig}
\usepackage{bm}
\usepackage[utf8]{inputenc}
\usepackage{graphics}
\usepackage{graphicx}
\usepackage{epsfig}
\usepackage{amssymb}
\usepackage{amsmath}
\usepackage{color}

\begin{document}
\title{Bound state solutions of the Klein Gordon equation with energy-dependent potentials}

\author{B.C. L\"{u}tf\"{u}o\u{g}lu}
\affiliation{Department of Physics,  Akdeniz University, 07058 Antalya, Turkey.}
\affiliation{Department of Physics,  University of Hradec Kr\'{a}lov\'{e}, Rokitansk\'{e}ho 62, 500\,03 Hradec Kr\'{a}lov\'{e}, Czechia.}

\author{A.N Ikot}
\affiliation{Department of Physics, Theoretical Physics Group, University of Port Harcourt, Choba, Port Harcourt, Nigeria.}

\author{M. Karakoc}
\affiliation{Department of Physics,  Akdeniz University, 07058 Antalya, Turkey.}

\author{G.T. Osobonye}
\affiliation{Department of Physics,Federal College of Education, Omuku, Rivers state, Nigeria.}

\author{A.T. Ngiangia}

\affiliation{Department of Physics, Theoretical Physics Group, University of Port Harcourt, Choba, Port Harcourt, Nigeria.}

\author{O. Bayrak}
\affiliation{Department of Physics,  Akdeniz University, 07058 Antalya, Turkey.}

\date{\today}
\begin{abstract}
In this manuscript, we investigate the exact bound state solution of the Klein-Gordon equation for an energy-dependent Coulomb-like vector plus scalar potential energies.  To the best of our knowledge, this problem is examined in literature with a constant and position dependent mass functions. As a novelty, we assume a mass-function that depends on energy and position and revisit the problem with the following cases: First, we examine the case where the mixed vector and scalar potential energy possess equal magnitude and equal sign as well as an opposite sign.  Then, we study pure scalar and pure vector cases.  In each case, we derive an analytic expression of the energy spectrum by employing the asymptotic iteration method. We obtain a non-trivial relation among the tuning parameters which lead the examined problem to a constant mass one.  Finally, we calculate the energy spectrum by the Secant method and show that the corresponding unnormalized wave functions satisfy the boundary conditions. We conclude the manuscript with a comparison of the calculated energy spectra versus tuning parameters.

\end{abstract}
\maketitle 

\section{Introduction}
One of the important goals in relativistic or non-relativistic quantum mechanics is to obtain an exact solution of a physical problem by employing potential energy \cite{Dirac_Book, Landau_Book, Greiner_Book}. It is found that among the analytically solvable potential energies, only a few of them are very appropriate to represent the physical systems. For example, Coulomb potential energy is used to describe the Hydrogen atom. Bound state solution of the Coulomb-like potential energy is investigated in details in the relativistic \cite{Greiner_Book} and non-relativistic \cite{Schiff_Book} equations.

Klein-Gordon (KG) equation is one of the relativistic equation that is introduced to describe the dynamics of the bosonic particles. Although it was first introduced about a century ago \cite{Klein_1926}, it still attracts the scientists' interest. Especially, researchers focus on the solutions of the KG equation with equal and unequal vector and scalar potential energies.  Among a large number of articles, we would like to underline the investigations that employ Woods-Saxon \cite{Lutfuoglu_et_al_2018_LK}, Hulth\'en \cite{dominguez_1989}, Morse \cite{Chen_2005}, Rosen-Morse \cite{Yi_et_al_2004} , Eckart \cite{Olgar_et_al_2006}, Manning-Rosen \cite{Jia_et_al_2013}, P\"oschl-Teller \cite{Chen_et_al_2009}, Kratzer \cite{Saad_et_al_2008}, Hylleraas \cite{Ikot_2012}, multiparameter \cite{Diao_et_al_2004}, exponential-type molecule \cite{Ikot_et_al_2016}, pseudoharmonic oscillator \cite{Ikhdair_et_al_2008},  Hartmann \cite{Chen_et_2006}, double ring-shaped oscillator \cite{Fa-lin_et_2004} potential energies.

In the last decades,  position-dependent mass (PDM) are gained popularity to interpret the properties of several quantum systems such as quantum liquids \cite{Arias_et_al_1994}, quantum wells \cite{Serra_et_al_1997}, quantum rings \cite{Li_et_al_2003}, quantum dots \cite{Peter_et_al_2008}, etc. More applications about quantum semiconductor structures can be found in the book of Weisbuch and  Vinter \cite{Weisbuch_Book}.

On the other hand, energy-dependent potential (EDP) energies are being used to describe the physical systems for a while \cite{Synder_1940, Schiff_1940, Green_1962}.  It is worth noting that these applications cover a wide area of physics. For example, in mathematical physics \cite{Znojil_2004, Znojil_et_al_2004, Gunther_et_al_2007, Schulze_2013}, quantum electronics \cite{Milanovic_et_al_1996, Belyavski_et_al_2007}, nuclear physics \cite{Niyogi_et_al_1981, McKellar_1983}, quantum chromodynamics \cite{Martynenko_2008, Sanctis_2009} etc... In recent times, the number of studies considering several EDP energies in both relativistic and non-relativistic wave equations is increased \cite{Formanek_et_al_2004, Lombard_et_al_2007, Martinez_2009, HH1, HH2, Gupta_et_al_2012, Ikot_Hassanabadi_2013, Benchikha_et_al_2013, Ikot_2, Benchikha_et_al_2014, Benchikha_et_al_2017_1, Benchikha_et_al_2017_2, Boumali_et_al_2017, Salti, Yesiltas_2, Boumali_et_al_2018, Lutfuoglu_Ikot_et_al_2019, Langueur_et_al_2019, Yesiltas_2019, Borrego_et_al_2020}.

In 2009, Ikhidair investigated the KG equation in $ 3+1$ dimension with PDM \cite{Ikhdair_2009}. Later, Hassanabadi \emph{et al.} explored the exact solutions of the D-dimensional KG equation with Coloumb-like EDP energy with a constant mass in 2012 \cite{HH2}. To the best of our knowledge, the KG equation with an energy PDM  under an EDP energy has not examined. Therefore, In this manuscript, we are motivated to investigate the bound state solutions of the Coulomb-like EDP energy by considering mixed vector and scalar potential energy in the presence of energy and PDM. We explore the bound state energy eigenvalues and corresponding wave functions for the KG particle in terms of following cases: mixed vector and scalar potential energies with equal magnitude and equal sign (EMES), mixed vector and scalar potential energies with equal magnitude and opposite sign (EMOS), pure vector (PV) potential energy, pure scalar (PS) potential energy and constant mass-energy limits.

We prepared the manuscript as follows. In sect. \ref{KG} we
introduced the radial KG equation with mixed vector and scalar
equation. Then, in subsection \ref{BSS}, we declared the position
and energy dependent mass function. Then, in its subsections, we
solved the radial KG equation by employing the AIM method in
different limits. In sect. \ref{WFS}, we obtained the wave function
solution by using the boundary conditions in terms of confluent
hypergeometric functions. Then, in sect. \ref{RD}, we employed the
numerical methods and calculated the energy spectrum and
corresponding wave functions for a neutral pion particle. Before we
concluded the manuscript in sect. \ref{Conc}, we analyzed the role
of the tuning parameters on the energy spectrum in the pure scalar
limits.

\section{Klein-Gordon equation} \label{KG}
In $(3+1)$ dimensions, in the presence of spherical symmetric vector, $V_v(r)$, and  scalar, $V_s(r)$, potential energies  the time independent KG equation with a PDM is given as \cite{Ikhdair_2009}
\begin{eqnarray}
\Bigg[\hbar^2c^2\nabla^2+\bigg(\Big(E-V_v(r)\Big)^2-\Big(m(r,E)c^2+V_s(r)\Big)^2\bigg)\Bigg]\phi(r)&=&0. \label{anadenklem0}
\end{eqnarray}
Here, $\hbar$ and $c$ represent the reduced Planck constant and the speed of light, respectively. $E$ is the energy of the bosonic particle. The mass term, $ m (r, E) $, depends on the energy in addition to the spatial coordinate. The decomposition of the spatial wave function into radial wave function, $R(r)$ and angular-dependent spherical harmonics, $Y_m^l(\theta,\varphi)$, ends up with
\begin{eqnarray}
\Bigg[\hbar^2c^2\bigg(\frac{d^2}{dr^2}-\frac{l(l+1)}{r^2}\bigg)+\bigg(\Big(E-V_v(r)\Big)^2-\Big(m(r,E)c^2+V_s(r)\Big)^2\bigg)\Bigg]u(r)&=&0, \label{anadenklem}\,\,\,\,\,\,
\end{eqnarray}
where $u(r) \equiv r R(r)$.

\subsection{Bound state solution} \label{BSS}
In this manuscript, we investigate bound state solutions of the KG equation with a position and energy dependent mass function that is defined by
\begin{eqnarray}
m(r,E)&=&m_0 \Bigg(1-\lambda b A \frac{(1+\delta E)}{r}\Bigg). \label{Massterm}
\end{eqnarray}
Here,  $\lambda$ and $m_0$ are the reduced Compton wavelength and the rest mass of the spin-0 particle, respectively. $b$ is the coupling constant. The term $ \frac {-A (1+\delta E)} {r} $ is a Coulomb-like energy-related potential energy function that will be called "vector potential energy" in the rest of the article.  $A$ represents the strength of the potential well, where $\delta$ determines the energy contribution. On the other hand, $b$ and $\delta$ can be seen as two tuning parameters that allow projecting the obtained analytic solutions on various physical problems. For instances, $(b=0)$, $(b=0, \,\, \delta=0)$, and $(b\neq0, \,\, \delta=0)$ cases represent the constant mass with energy dependent Coulomb-like potential energy, Coulomb-like potential, and the position dependent  as well as energy independent solutions, respectively.

Note that, we can express the investigated mass energy term as follows:
\begin{eqnarray}
m(r,E)c^2 &=& m_0c^2+b \hbar V_v(r,E). \label{varmass}
\end{eqnarray}

\subsubsection{Mixed vector and scalar potential energies with equal magnitude and equal sign case}

In this subsection, we use attractive scalar potential energy that has EMES with the vector potential energy.
\begin{eqnarray}
V(r,E)\equiv V_v(r,E) = V_s(r,E)= -\frac {A (1+\delta E)} {r} \label{potenEMES}
\end{eqnarray}
The EMES potential energies condition turns Eq. (\ref{anadenklem}) into
\begin{eqnarray}
\Bigg[\Bigg(\frac{d^2}{dr^2}-\frac{l(l+1)}{r^2}\Bigg)+\frac{1}{\hbar^2c^2}\Bigg(E^2-m(r,E)^2c^4-2V(r,E)\Big(E+m(r,E)c^2\Big)\Bigg)\Bigg]u(r)&=&0. \,\,\,\,\,\,\,\, \label{anadenklem1}
\end{eqnarray}
Then, we employ Eqs. (\ref{Massterm}) and (\ref{potenEMES}) in Eq. (\ref{anadenklem1}) and we obtain
\begin{eqnarray}
&&u''(r)+\Bigg[\frac{E^2-m_0^2c^4}{\hbar^2c^2}+ \Bigg(\frac{2 \lambda b A m_0^2c^4}{\hbar^2c^2}+\frac{2A(E+m_0c^2)}{\hbar^2c^2}\Bigg) \frac{(1+\delta E)}{r} \nonumber \\
&&-\Bigg(\frac{m_0^2c^4\lambda^2 b^2 A^2 }{\hbar^2c^2}+\frac{2\lambda b A^2 m_0c^2}{\hbar^2c^2}+\frac{l(l+1)}{(1+\delta E)^2}\Bigg)\frac{(1+\delta E)^2}{r^2}\Bigg]u(r)=0.
\end{eqnarray}
We introduce a new coordinate transformation,
$z=(1+\delta E)r$, hence, $u(r)\rightarrow v(z)$. We get
\begin{eqnarray}
&&\frac{d^2v}{dz^2}+\Bigg(-\tau^2+\frac{\beta^2 }{z}-\frac{\eta (\eta+1)}{z^2}\Bigg)v=0. \label{denklem1}
\end{eqnarray}
where
\begin{eqnarray}
\tau^2&\equiv&-\frac{E^2-m_0^2c^4}{\hbar^2c^2(1+\delta E)^2}, \\
\beta^2&\equiv& \frac{2A\big(E+m_0c^2+\lambda b m_0^2c^4\big)}{\hbar^2 c^2}, \\
\eta (\eta+1)&\equiv&\Bigg(\frac{m_0^2c^4\lambda^2 b^2 A^2 }{\hbar^2c^2}+\frac{2\lambda b A^2 m_0c^2}{\hbar^2c^2}+\frac{l(l+1)}{(1+\delta E)^2}\Bigg)(1+\delta E)^2.
\end{eqnarray}
We put forward an ansatz by examine the asymptotic behaviours
\begin{eqnarray}
v(z)&=& e^{-\tau z} z^{\eta+1} f(z) \label{denklem2}
\end{eqnarray}
and we find
\begin{eqnarray}
f''(z)&=&\frac{2\big[\tau z-\big(\eta+1\big)\big]}{z}f'(z)+\frac{2\tau\big(\eta+1\big)- \beta^2}{z}f(z). \label{AIMicin1}
\end{eqnarray}

In order to obtain the energy eigenvalues, we employ the Asymptotic
Iteration Method (AIM) which can be found in details in
Ref.\cite{Ciftci_et_al_2003}. According to AIM, we firstly find the
homogeneous linear second-order differential equation of the form
$\varphi''(z)=\lambda_0(z)\varphi'(z)+s_0(z) \varphi(z)$. Comparing
this equation with Eq. (\ref{AIMicin1}), we get 
\begin{eqnarray}
\lambda_0(z)&=& \frac{2\big[\tau z-\big(\eta+1\big)\big]}{z}, \\
s_0(z)&=& \frac{2\tau\big(\eta+1\big)- \beta^2}{z}.
\end{eqnarray}
By using the iteration formula
$\lambda_{n}(z)=\lambda_{n-1}'(z)+\lambda_0(x)\lambda_{n-1}(z)+s_{n-1}(z)$
and $s_{n}(z)=s_{n-1}'(z)+s_0(z)\lambda_{n-1}(z)$ as well as the
quantization condition
$\frac{s_n(z)}{\lambda_n(z)}=\frac{s_{n-1}(z)}{\lambda_{n-1}(z)}$,
we obtain,
\begin{eqnarray}
\frac{s_0}{\lambda_0}&=&\frac{s_1}{\lambda_1}  \Rightarrow \tau_0= \frac{\beta^2}{2(\eta_0+1)},\,\,\,\, n=0. \\
\frac{s_1}{\lambda_1}&=&\frac{s_2}{\lambda_2}  \Rightarrow \tau_1= \frac{\beta^2}{2(\eta_1+2)},\,\,\,\, n=1. \\
\frac{s_2}{\lambda_2}&=&\frac{s_3}{\lambda_3}  \Rightarrow \tau_2= \frac{\beta^2}{2(\eta_2+3)},\,\,\,\, n=2.
\end{eqnarray}
We can easily found a general formula for a given $n$ quantum number
as follows,
\begin{eqnarray}\label{quanti}
\tau_n&=&\frac{\beta^2}{2\big(\eta_n+n+1\big)}.
\end{eqnarray}
We substitute $\tau$, $\beta$ and $\eta$ parameters into Eq.
(\ref{quanti}) and find
\begin{eqnarray}
\sqrt{\frac{m_0^2c^4-E_{nl}^2}{(1+\delta E_{nl})^2}}&=&\frac{\frac{A}{\hbar c}\big(E_{nl}+m_0c^2+\lambda b m_0^2c^4\big)}{n+\frac{1}{2}\mp \sqrt{\frac{1}{4}+\Bigg(\frac{m_0^2c^4\lambda^2 b^2 A^2 }{\hbar^2c^2}+\frac{2\lambda b A^2 m_0c^2}{\hbar^2c^2}+\frac{l(l+1)}{(1+\delta E_{nl})^2}\Bigg)(1+\delta E_{nl})^2}} \label{QuantizationEMES}
\end{eqnarray}
where $n=0,1,2,3,\cdots$.

\subsubsection{Mixed vector and scalar potential energies with equal magnitude and opposite sign case}

In this subsection, we use repulsive scalar potential energy that possesses equal magnitude with the vector potential energy but with an opposite sign.
\begin{eqnarray}
V(r,E)\equiv V_v(r,E) = -V_s(r,E)= -\frac {A (1+\delta E)} {r} \label{potenEMOS}
\end{eqnarray}
We employ the EMOS condition in Eq. (\ref{anadenklem}) and we derive
\begin{eqnarray}
\Bigg[\Bigg(\frac{d^2}{dr^2}-\frac{l(l+1)}{r^2}\Bigg)+\frac{1}{\hbar^2c^2}\Bigg(E^2-m(r,E)^2c^4-2V(r,E)\Big(E-m(r,E)c^2\Big)\Bigg)\Bigg]u(r)&=&0. \,\,\,\,\,\,\,\, \label{anadenklem2}
\end{eqnarray}
We follow similar steps and find Eq. (\ref{denklem1}) with different parameters
\begin{eqnarray}
\tau^2&\equiv&-\frac{E^2-m_0^2c^4}{\hbar^2c^2(1+\delta E)^2}, \\
\beta^2&\equiv& \frac{2A\big(E-m_0c^2+\lambda b m_0^2c^4\big)}{\hbar^2 c^2}, \\
\eta (\eta+1)&\equiv&\Bigg(\frac{m_0^2c^4\lambda^2 b^2 A^2 }{\hbar^2c^2}-\frac{2\lambda b A^2 m_0c^2}{\hbar^2c^2}+\frac{l(l+1)}{(1+\delta E)^2}\Bigg)(1+\delta E)^2.
\end{eqnarray}
First, we employ the same ansatz and then, apply the AIM method, we get the quantization condition as follows
\begin{eqnarray}
\sqrt{\frac{m_0^2c^4-E_{nl}^2}{(1+\delta E_{nl})^2}}&=&\frac{\frac{A}{\hbar c}\big(E_{nl}-m_0c^2+\lambda b m_0^2c^4\big)}{n+\frac{1}{2}\mp \sqrt{\frac{1}{4}+\Bigg(\frac{m_0^2c^4\lambda^2 b^2 A^2 }{\hbar^2c^2}-\frac{2\lambda b A^2 m_0c^2}{\hbar^2c^2}+\frac{l(l+1)}{(1+\delta E_{nl})^2}\Bigg)(1+\delta E_{nl})^2}}  \label{QuantizationEMOS}
\end{eqnarray}
where $n$ is an integer.

\subsubsection{Pure vector potential energy case}
In this subsection, we examine the case where the scalar potential energy is equal to zero.
\begin{eqnarray}
V_s(r)&=&0, \\
V(r,E)\equiv V_v(r,E)& =& -\frac {A(1+\delta E)}{r} \label{potenPV}
\end{eqnarray}
We would like to remind you that since the variable mass has vector potential energy coupling, even in the absence of scalar potential energy, the KG equation is not in the minimal coupling form. We use the pure vector condition in Eq. (\ref{anadenklem1}) and we find
\begin{eqnarray}
\Bigg[\Bigg(\frac{d^2}{dr^2}-\frac{l(l+1)}{r^2}\Bigg)+\frac{1}{\hbar^2c^2}\Bigg(E^2-m(r,E)^2c^4-V(r,E)\Big(2E-V(r,E)\Big)\Bigg)\Bigg]u(r)&=&0. \,\,\,\,\,\,\,\, \label{purevectoranademklem}
\end{eqnarray}
We go through the same procedure given in details above and we obtain Eq. (\ref{denklem1}) with different parameters as
\begin{eqnarray}
\tau^2&\equiv&-\frac{E^2-m_0^2c^4}{\hbar^2c^2(1+\delta E)^2}, \\
\beta^2&=& \frac{2A\big(E+\lambda b m_0^2c^4\big)}{(\hbar c)^2}, \\
\eta (\eta+1)&=&\Bigg(\frac{A^2}{\hbar^2c^2}\big(m_0^2c^4\lambda^2 b^2-1\big)+\frac{l(l+1)}{(1+\delta E)^2}\Bigg)(1+\delta E)^2.
\end{eqnarray}
After, we use the same ansatz and the several iterations in the AIM method, we derive the quantization condition as given
\begin{eqnarray}
\sqrt{\frac{m_0^2c^4-E_{nl}^2}{(1+\delta E_{nl})^2}}&=&\frac{\frac{A}{\hbar c}\big(E_{nl}+\lambda b m_0^2c^4\big)}{n+\frac{1}{2}\mp \sqrt{\frac{1}{4}+\Bigg(\frac{A^2}{\hbar^2c^2}\big(m_0^2c^4\lambda^2 b^2-1\big)+\frac{l(l+1)}{(1+\delta E_{nl})^2}\Bigg)(1+\delta E_{nl})^2}} \label{QuantizationPV}
\end{eqnarray}
Here, the quantum number $n$ is an integer.

\subsubsection{Pure scalar potential energy case}
In this subsection, we investigate the case where the vector potential energy is equal to zero. Consequently, the mass term does not depend on either position or energy. On the other hand, the mass term couples with an attractive scalar potential energy.
\begin{eqnarray}
V_v(r)&=&0, \\
V(r,E)\equiv V_s(r,E)& =& -\frac {A(1+\delta E)}{r}. \label{potenPS}
\end{eqnarray}
In this case the KG equation given in Eq. (\ref{anadenklem}) turns to be
\begin{eqnarray}
\Bigg[\Bigg(\frac{d^2}{dr^2}-\frac{l(l+1)}{r^2}\Bigg)+\frac{1}{\hbar^2c^2}\Bigg(E^2-m_0^2c^4-V(r,E)\Big(2m_0c^2+V(r,E)\Big)\Bigg)\Bigg]u(r)&=&0. \,\,\,\,\,\,\,\, \label{purescalaranademklem}
\end{eqnarray}
After basic algebra which is given in details in the previous subsections, we find the quantization condition
as follows:
\begin{eqnarray}
\sqrt{\frac{m_0^2c^4-E_{nl}^2}{(1+\delta E_{nl})^2}}&=&\frac{\frac{A}{\hbar c} m_0 c^2\big(1+\lambda b m_0c^2\big)}{n+\frac{1}{2}\mp \sqrt{\frac{1}{4}+\Bigg(\frac{A^2}{\hbar^2c^2}\big(1+m_0 c^2\lambda b\big)^2+\frac{l(l+1)}{(1+\delta E_{nl})^2}\Bigg)(1+\delta E_{nl})^2}} \label{QuantizationPS}
\end{eqnarray}

\subsubsection{Constant mass energy limit}
In this section, we discuss a limit where the vector and scalar potential energy couplings to the mass term vanish. Basically this limit is obtained by employing the definition of the non constant mass  term, which is given in Eq. (\ref{varmass}), in the KG equation, which is expressed in Eq. (\ref{anadenklem}). We find
\begin{eqnarray}
\Bigg[\hbar^2c^2\bigg(\frac{d^2}{dr^2}-\frac{l(l+1)}{r^2}\bigg)+\bigg(\Big(E-V_v(r,E)\Big)^2-\Big( m_0c^2+b \hbar V_v(r,E)+V_s(r,E)\Big)^2\bigg)\Bigg]u(r)&=&0. \,\,\,\, \label{ConsMEL}\,\,\,\,\,\,
\end{eqnarray}
Since $b$ is an arbitrary parameter, it is always possible to obtain
\begin{eqnarray}
 \Big(m_0c^2+b \hbar V_v(r,E)+V_s(r,E)\Big)^2 = m_0^2c^4.
\end{eqnarray}
via $b=-\frac{V_s(r,E)}{\hbar V_v(r,E)}$. For instance if we assume an attractive vector  and a repulsive scalar potential energies with different magnitudes
\begin{eqnarray}
 V_v(r,E)&=&-\frac{A}{r}(1+\delta E), \\
 V_s(r,E)&=& \frac{B}{r}(1+\delta E)
\end{eqnarray}
with position and energy dependence. Then, we find
 \begin{eqnarray}
b=\frac{B}{ \hbar A}
 \end{eqnarray}
Note that, the tuning parameter $b$ does not need to be a positive number.

Before we conclude this subsection, we strongly advice to read papers that examine some classes of exactly-solvable KG equations within $V_v(r)\equiv V_0 + \beta V_s(r)$ relation where $V_0$ and $\beta$ are arbitrary constants \cite{Chen_2005, Chen_et_al_2006, Dutra_et_al_2006}.

\subsection{Wave function solution} \label{WFS}
In this section we obtain the wave function solution. We start with a new transformation, $y=2\tau z$, in
Eq. (\ref{AIMicin1}). We find
\begin{eqnarray}
\frac{d^2g(y)}{dy^2}+\frac{2\big(\eta+1\big)-y}{y}\frac{dg(y)}{dy}+\frac{\frac{\beta^2}{2 \tau}-\big(\eta+1\big)}{ y}g(y)&=&0.\label{bir}
\end{eqnarray}
This equation is similar to the confluent hypergeometric equation \cite{Abramowitz_et_al_Book}
\begin{eqnarray}
xy''+(c-x)y'-ay&=&0 \label{iki}
\end{eqnarray}
which has the solution
\begin{eqnarray}
y&=&N_1 \,_1F_1(a,c,x)+N_2 \,_1U_1(a,c,x)
\end{eqnarray}
Here, $\,_1F_1(a,c,x)$ and $\,_1U_1(a,c,x)$ are the first and second kind confluent hypergeometric functions, respectively. We obtain the coefficients with the match of the Eq. (\ref{bir}) and Eq. (\ref{iki})
\begin{eqnarray}
a&=&\frac{2\tau(\eta+1)-\beta^2}{2\tau} \\
c&=&2(\eta+1).
\end{eqnarray}
and therefore the solution of Eq. (\ref{AIMicin1})
\begin{eqnarray}
g(r)&=&N_1 \,_1F_1\Bigg(\frac{2\tau(\eta+1)-\beta^2}{2\tau},2(\eta+1),2\tau(1+\delta E)r\Bigg)\nonumber \\&+&N_2 \,_1U_1\Bigg(\frac{2\tau(\eta+1)-\beta^2}{2\tau},2(\eta+1),2\tau(1+\delta E)r\Bigg)
\end{eqnarray}
Then the wave function $u(r)$ is found to be
\begin{eqnarray}
u(r)&=&N_1 e^{-\tau (1+\delta E) r }\big((1+\delta E)r\big)^{\eta+1} \,_1F_1\Bigg(\frac{2\tau(\eta+1)-\beta^2}{2\tau},2(\eta+1),2\tau(1+\delta E)r\Bigg).\,\,\,\,
\end{eqnarray}
Note that second kind confluent hypergeometric function does not satisfy the boundary conditions. Hence $N_2$ is chosen to be zero.

\section{Results and Discussions} \label{RD}
In this section, in order to get quantitative analysis, we
illustrate the energy eigenvalues of a neutral pion. The rest mass
energy and Compton wavelength of the neutral pion are $134.977$
$MeV$ and $1.462$ $fm$, respectively. We assume the strength of the
potential energy parameter, $A$, as $200$ $MeV \cdot fm$. We assign
negative and positive values to the tuning parameters $\delta$ and
$b$ in addition to their vanishing values. First, we use Eq.
(\ref{QuantizationEMES}) to calculate the energy spectra in the EMES
limit. We tabulate the obtained energy spectra in Table
\ref{EMEStable}. We plot the unnormalized wave functions, the vector
potential energy and the mass function in Fig. \ref{EMESwaves}. Note
that, when $b$ is zero, $\lambda b$ becomes zero and the mass loses
both energy and position dependence. However, energy dependency on
the potential energy continues. To comprehend the effect of energy
dependence, we use three values in which the tuning parameter
$\delta$ has negative,  $-0.003$ $MeV^{-1}$, and positive $0.003$
$MeV^{-1}$, values in addition to zero. In this case, the spectra
consist only of the upper or lower energy eigenvalues. During the
increase of delta parameter from the negative value to the positive
value, we observe that lower energy eigenvalues increase and upper
energy eigenvalues decrease. When $b$ has negative value, we realize
that lower energy eigenvalues do not occur. In this case, the
spectra consist only of the upper energy eigenvalues, and as the
delta increases, the values of the energy eigenvalues decrease. This
decrease is relatively smaller as quantum numbers increase. When $b$
has a positive value, the upper and lower energy eigenvalues exist
in the energy spectra in Table \ref{EMEStable}.

Then, we examine the energy spectra of the EMOS limit by solving Eq.
(\ref{QuantizationEMOS}). We find that the energy spectra persist
non zero values only in positive values of the tuning parameter $b$.
We tabulate the calculated energy spectra in Table \ref{EMOStable}.
One of the authors of this study, BCL, in a study published in 2018,
examined the generalized symmetric Woods-Saxon potential energy with
a constant mass term in the KG equation and showed that an energy
spectrum at the EMOS limit cannot be obtained \cite{Lutfuoglu_2018}.
The results obtained in this study are in agreement with the results
given by BCL. In addition to those results, we show that a spectrum
can be calculated with upper energy eigenvalues for a specific value
of the effective mass.

Then, we calculate the energy spectra in the PV limit, where
$V_s(r)=0$. We present the energy spectra in Tab. \ref{PVtable}.  We
observe that the ground state eigenvalue does not exist, instead the
lowest eigenvalue is $E_{11}$. In all cases, the eigenvalues have
positive values. The increase of $\delta$ parameter with the
negative value of $b$ parameter creates a decrease in the value of
the lowest eigenvalue. Furthermore, these decreases have an
increment when $b$ becomes positive.

We study the pure scalar limit and tabulate the energy spectra in
Tab. \ref{PStable}. In each particular sub-case, two energy
eigenvalues, namely upper and lower eigenvalues are obtained.
Similar to the pure vector case, a decrease between the energy
eigenvalues is observed in the increasing values of $\delta$ for a
fixed $b$ value. To demonstrate it, we plot  Fig. \ref{delta}.
There, we illustrate the variation of particular upper and lower
energy eigenvalues of the spectra via $\delta$ parameter for the
fixed negative value of $b$. When the potential energy is
independent of the energy and the mass has a constant value, the
energy eigenvalues become symmetrical as expected. In addition, when
the mass and potential energy are independent of energy, a symmetry
of upper and lower energy eigenvalues exist. Finally, in Fig.
\ref{lamdab}, we demonstrate the upper and lower energy eigenvalues
of the spectra versus $\lambda b$ values for a negative constant
value of the tuning parameter $\delta$. We see a higher
differentiation of the eigenvalues for lower quantum numbers. We
observe a decrease in upper and an increase in lower eigenvalues for
small quantum numbers. As the quantum numbers for a fixed $\lambda
b$ increase, the difference between the two adjacent energy levels
decreases.

\section{Conclusion} \label{Conc}
In this paper, we obtained the bound state solution of a Klein
Gordon particle whose mass varies via position and energy.  We took
into account an energy-dependent Coulomb-like potential energy and
employed the asymptotic iteration method.  We examined the solution
in five different limits, i.e. mixed vector and scalar potential
energies with equal magnitudes with equal/opposite signs, pure
vector, pure scalar,  and constant mass.  Except for the trivial
constant mass limit, in each cases, we derived a transcendental
equation to calculate the energy spectrum. We showed that the radial
wave functions are occurring in terms of the confluent
hypergeometric functions. Next, we considered a neutral pion
particle and used the Secant numerical methods to obtain an energy
spectrum in each limit. Although the potential energy parameters
were chosen at random, we have set various tuning parameters to
investigate all critical situations. We verified the tabulated
energy eigenvalues by examining their corresponding wave function
behavior which fulfills the boundary conditions. Finally, we
discussed the role of the tuning parameters in the pure scalar
limit. We believe that the results obtained in this article would
have applications in physics.

\center{

\newpage
\begin{figure}[hbtp!]
\includegraphics[width=0.7\textwidth,clip=true]{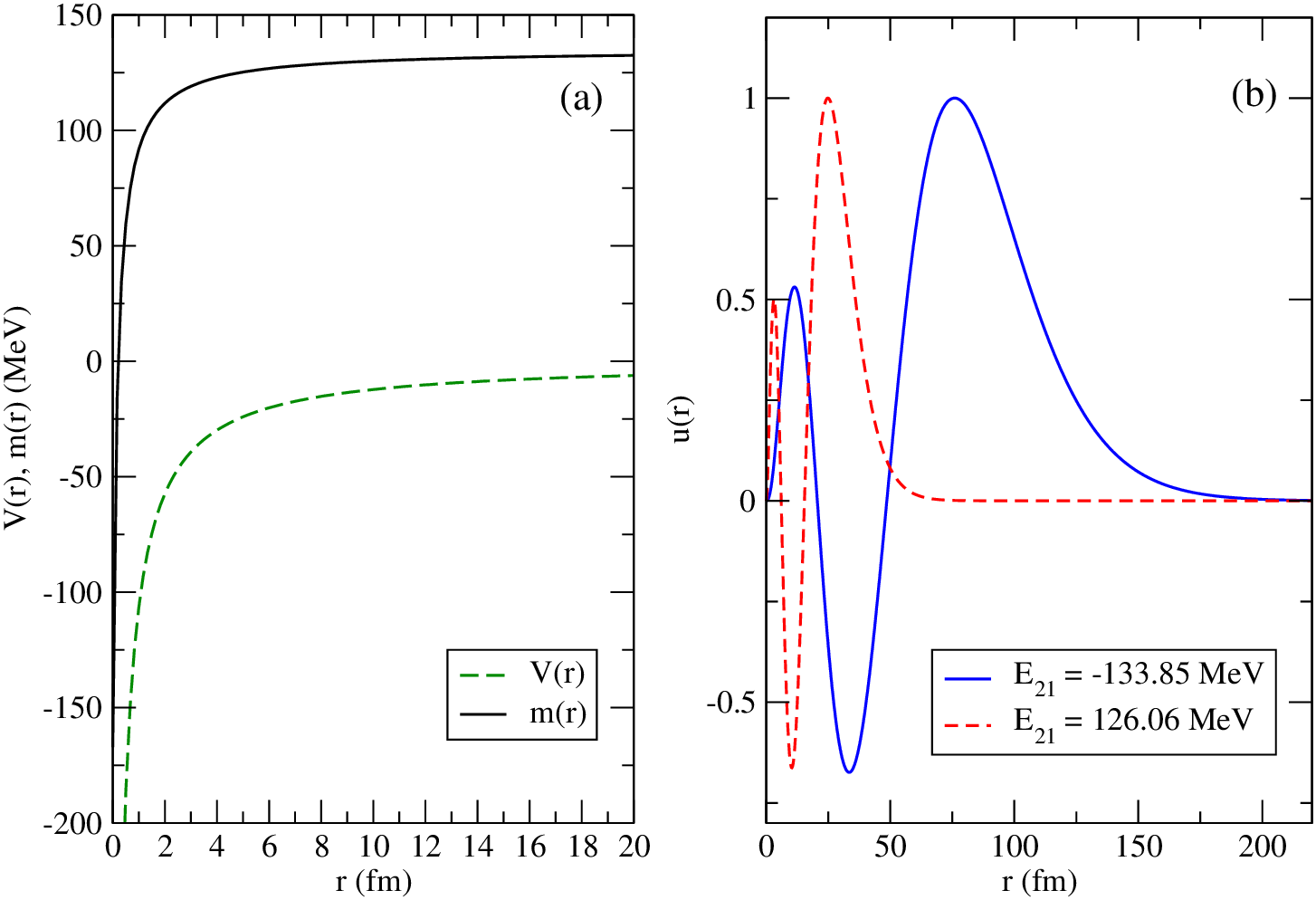}
\caption{(a) The potential energy $V(r)$ and mass $m(r)$ as a
function of radius. (b) Unnormalized wave functions for the positive
and negative energy eigenvalues $E_{nl}$ in the EMES limit.}
\label{EMESwaves}
\end{figure}

\newpage
\begin{figure}[hbtp!]
\includegraphics[width=0.75\textwidth,clip=true]{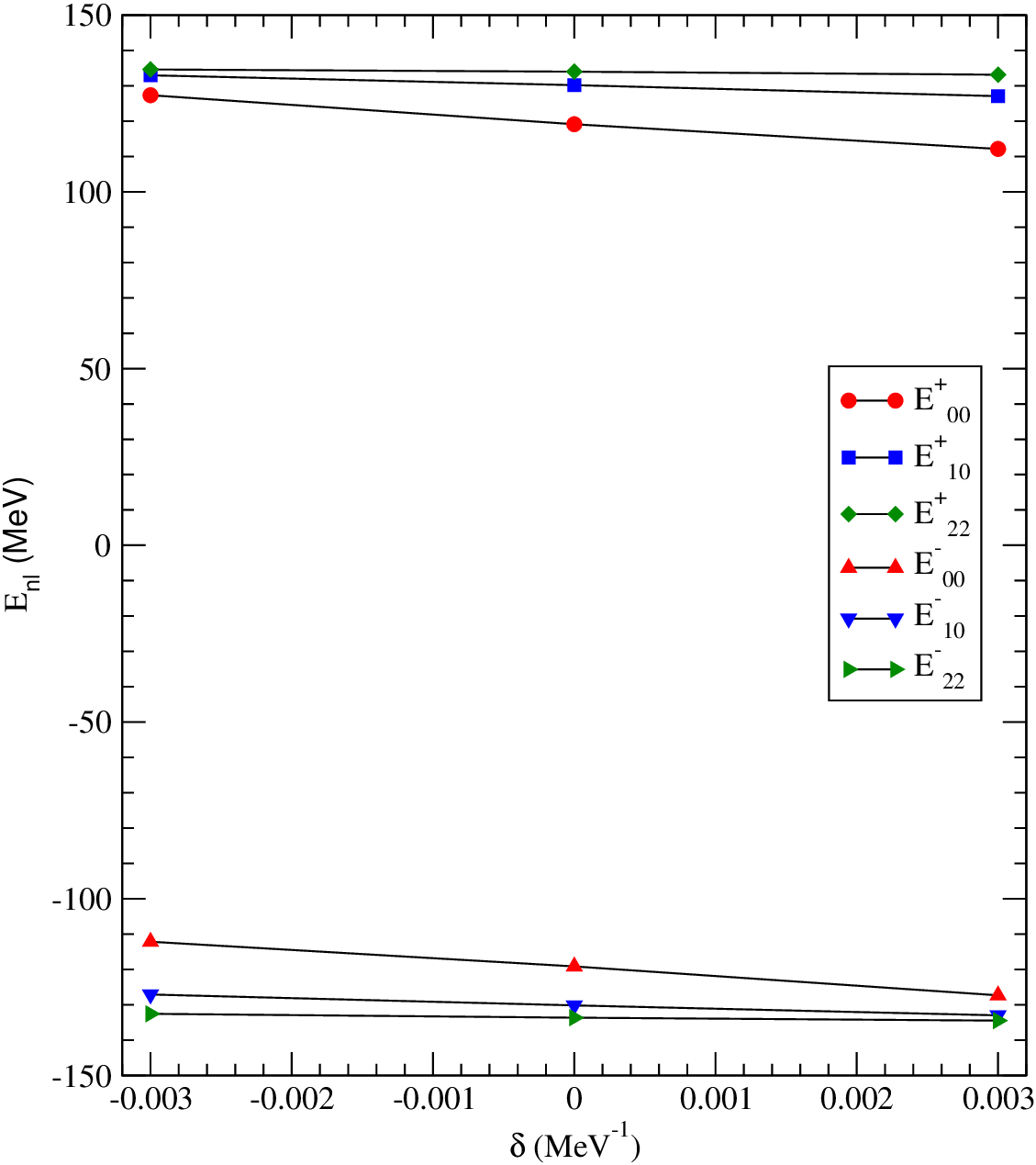}
\caption{The differentiation of particular upper and lower energy eigenvalues versus $\delta$ in the pure scalar case. Here, the other tuning parameter $\lambda b=-0.003$ $MeV^{-1}$.}
\label{delta}
\end{figure}

\newpage
\begin{figure}[hbtp!]
\includegraphics[width=0.75\textwidth,clip=true]{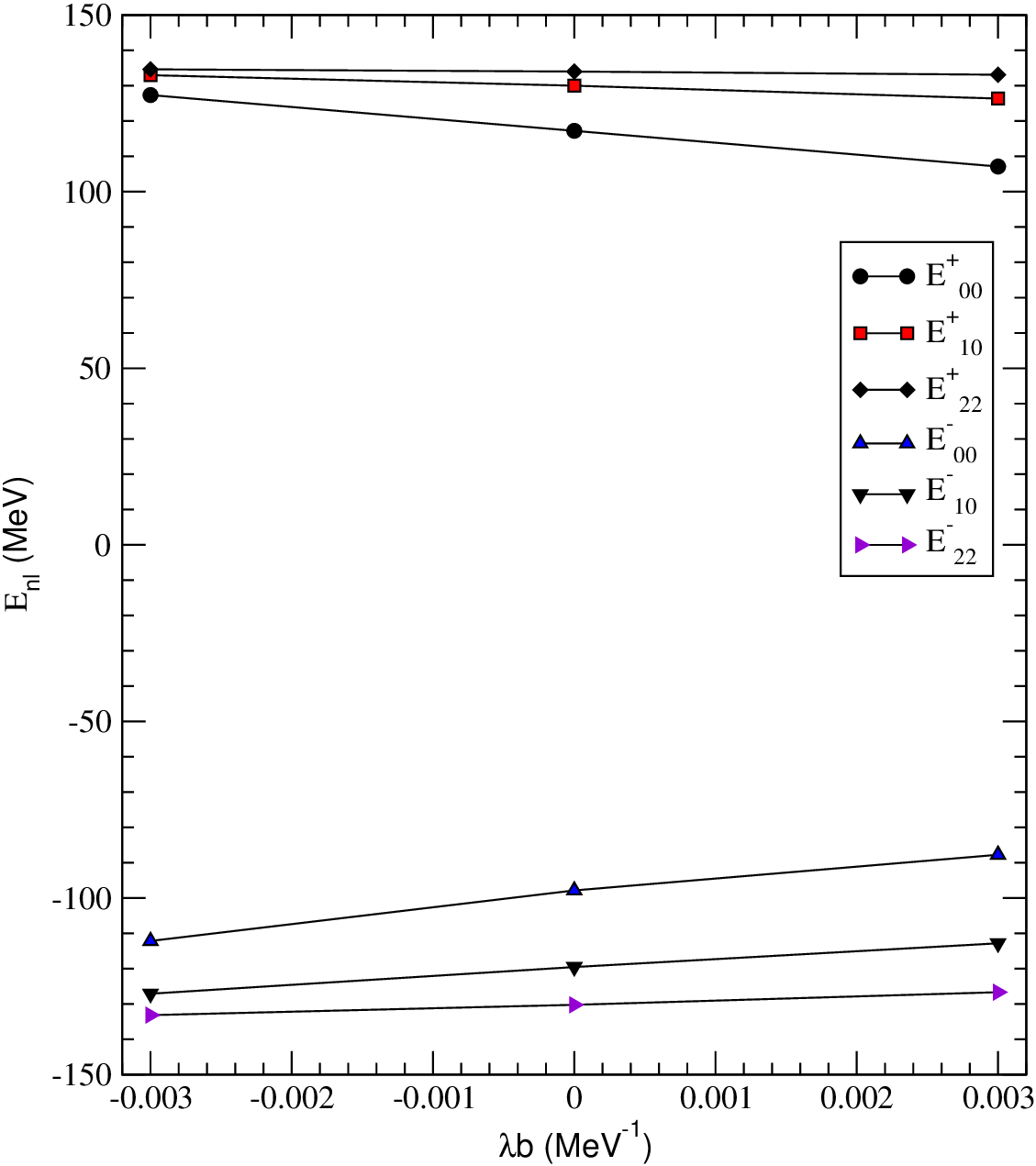}
\caption{The differentiation of particular upper and lower energy eigenvalues versus $\lambda b$ in the pure scalar case. Here, the other tuning parameter $\delta=-0.003$ $MeV^{-1}$.}
\label{lamdab}
\end{figure}


\newpage
\begin{sidewaystable}
\begin{tabular}{rrrrrrrrrrrr}
\hline
  $\delta$   & $\lambda b$  &   $E_{00}$   &   $E_{10}$   &   $E_{11}$   &   $E_{20}$   &   $E_{21}$   &   $E_{22}$   &   $E_{30}$   &   $E_{31}$   &   $E_{32}$   &   $E_{33}$   \\
\hline
    -0.00300 &     -0.00300 &     None     &     None     &     None     &     None     &     None     &     None     &     None     &     None     &     None     &     None     \\
             &              &     None     &     None     &    127.37813 &     None     &    130.83021 &    132.40021 &    129.92379 &    132.36407 &    133.19980 &    133.68234 \\
    -0.00300 &      0.00000 &     -3.03943 &     None     &     None     &     None     &     None     &     None     &     None     &     None     &     None     &     None     \\
             &              &     None     &    106.22197 &    123.27091 &    123.27091 &    128.59284 &    130.94895 &    128.59284 &    130.94895 &    132.20124 &    132.94711 \\
    -0.00300 &      0.00300 &   -127.55222 &   -132.19801 &   -133.07406 &   -133.48904 &   -133.85027 &   -134.17528 &   -134.04496 &   -134.22986 &   -134.41219 &   -134.54598 \\
             &              &     29.88598 &    103.11331 &    118.71065 &    120.60345 &    126.05962 &    129.21955 &    126.80482 &    129.32736 &    131.01021 &    132.05808 \\
     0.00000 &     -0.00300 &     None     &     None     &     None     &     None     &     None     &     None     &     None     &     None     &     None     &     None     \\
             &              &     None     &     None     &    114.45942 &     None     &    123.30931 &    127.81796 &     None     &    127.51480 &    129.98951 &    131.35124 \\
     0.00000 &      0.00000 &     -1.81465 &     None     &     None     &     None     &     None     &     None     &     None     &     None     &     None     &     None     \\
             &              &     None     &     79.81538 &    107.32122 &    107.32122 &    118.69067 &    124.32234 &    118.69067 &    124.32234 &    127.48761 &    129.43379 \\
     0.00000 &      0.00300 &   -129.64146 &   -133.19720 &   -133.88866 &   -134.07683 &   -134.34733 &   -134.54808 &   -134.43223 &   -134.56603 &   -134.67698 &   -134.75200 \\
             &              &     22.77861 &     83.76911 &    101.24833 &    106.48183 &    114.41062 &    120.62914 &    117.00422 &    121.19953 &    124.78258 &    127.26083 \\
     0.00300 &     -0.00300 &     None     &     None     &     None     &     None     &     None     &     None     &     None     &     None     &     None     &     None     \\
             &              &     None     &     None     &     98.91588 &     None     &    112.76280 &    121.45996 &     None     &    120.24283 &    125.36472 &    127.98363 \\
     0.00300 &      0.00000 &     -1.29096 &     None     &     None     &     None     &     None     &     None     &     None     &     None     &     None     &     None     \\
             &              &     None     &     63.05730 &     92.57467 &     92.57467 &    107.70124 &    116.20530 &    107.70124 &    116.20530 &    121.36605 &    124.69813 \\
     0.00300 &      0.00300 &   -132.17136 &   -134.17013 &   -134.55301 &   -134.59518 &   -134.73627 &   -134.81915 &   -134.75456 &   -134.82187 &   -134.86714 &     None     \\
             &              &     18.73778 &     71.69401 &     87.71693 &     95.21508 &    103.47799 &    111.34457 &    107.83252 &    112.63070 &    117.52188 &    121.28761\\
\hline
\end{tabular}
\caption{Calculated energy spectrum in the EMES limit of the mixed potential energies. Energy eigenvalues are denoted with two quantum numbers $n$ and $l$ as $E_{nl}$. The units of  $\delta$ and $\lambda b$ are $MeV^{-1}$. The energy eigenvalue is in  $MeV$ unit.}
\label{EMEStable}
\end{sidewaystable}

\newpage
\begin{sidewaystable}
\begin{tabular}{rrrrrrrrrrrr}
\hline
  $\delta$   & $\lambda b$  &   $E_{00}$   &   $E_{10}$   &   $E_{11}$   &   $E_{20}$   &   $E_{21}$   &   $E_{22}$   &   $E_{30}$   &   $E_{31}$   &   $E_{32}$   &   $E_{33}$   \\
\hline
    -0.00300 &     -0.00300 &     None     &     None     &     None     &     None     &     None     &     None     &     None     &     None     &     None     &     None     \\
             &              &     None     &     None     &     None     &     None     &     None     &     None     &     None     &     None     &     None     &     None     \\
    -0.00300 &      0.00000 &     None     &     None     &     None     &     None     &     None     &     None     &     None     &     None     &     None     &     None     \\
             &              &     None     &     None     &     None     &     None     &     None     &     None     &     None     &     None     &     None     &     None     \\
    -0.00300 &      0.00300 &     None     &     None     &     None     &     None     &     None     &     None     &     None     &     None     &     None     &     None     \\
             &              &     None     &    133.47228 &    134.50966 &    134.39701 &    134.71651 &    134.81349 &    134.67191 &    134.81129 &    134.86369 &    134.89421 \\
     0.00000 &     -0.00300 &     None     &     None     &     None     &     None     &     None     &     None     &     None     &     None     &     None     &     None     \\
             &              &     None     &     None     &     None     &     None     &     None     &     None     &     None     &     None     &     None     &     None     \\
     0.00000 &      0.00000 &     None     &     None     &     None     &     None     &     None     &     None     &     None     &     None     &     None     &     None     \\
             &              &     None     &     None     &     None     &     None     &     None     &     None     &     None     &     None     &     None     &     None     \\
     0.00000 &      0.00300 &     None     &     None     &     None     &     None     &     None     &     None     &     None     &     None     &     None     &     None     \\
             &              &     None     &     None     &    133.55348 &     None     &    134.19336 &    134.50388 &     None     &    134.48332 &    134.64992 &    134.74036 \\
     0.00300 &     -0.00300 &     None     &     None     &     None     &     None     &     None     &     None     &     None     &     None     &     None     &     None     \\
             &              &     None     &     None     &     None     &     None     &     None     &     None     &     None     &     None     &     None     &     None     \\
     0.00300 &      0.00000 &     None     &     None     &     None     &     None     &     None     &     None     &     None     &     None     &     None     &     None     \\
             &              &     None     &     None     &     None     &     None     &     None     &     None     &     None     &     None     &     None     &     None     \\
     0.00300 &      0.00300 &     None     &     None     &     None     &     None     &     None     &     None     &     None     &     None     &     None     &     None     \\
             &              &     None     &     None     &    131.74999 &     None     &    133.23676 &    134.00740 &     None     &    133.89988 &    134.30896 &    134.50141\\
\hline
\end{tabular}
\caption{Calculated energy spectrum in the EMOS limit of the mixed potential energies. Energy eigenvalues are denoted with two quantum numbers $n$ and $l$ as $E_{nl}$. The units of  $\delta$ and $\lambda b$ are $MeV^{-1}$. The energy eigenvalue is in  $MeV$ unit.}
\label{EMOStable}
\end{sidewaystable}

\newpage
\begin{sidewaystable}[tbp]
\begin{tabular}{rrrrrrrrrrrr}
\hline
  $\delta$   & $\lambda b$  &   $E_{00}$   &   $E_{10}$   &   $E_{11}$   &   $E_{20}$   &   $E_{21}$   &   $E_{22}$   &   $E_{30}$   &   $E_{31}$   &   $E_{32}$   &   $E_{33}$   \\
\hline
    -0.00300 &     -0.00300 &     None     &     None     &     None     &     None     &     None     &     None     &     None     &     None     &     None     &     None     \\
             &              &     None     &     None     &    133.95087 &     None     &    134.40744 &    134.62193 &     None     &    134.61559 &    134.73115 &    134.79771 \\
    -0.00300 &      0.00000 &     None     &     None     &     None     &     None     &     None     &     None     &     None     &     None     &     None     &     None     \\
             &              &     None     &     None     &    132.00638 &     None     &    133.34195 &    133.96617 &     None     &    133.94393 &    134.27837 &    134.46865 \\
    -0.00300 &      0.00300 &     None     &     None     &     None     &     None     &     None     &     None     &     None     &     None     &     None     &     None     \\
             &              &     None     &     None     &    129.06586 &     None     &    131.74488 &    132.97652 &     None     &    132.93977 &    133.59648 &    133.97227 \\
     0.00000 &     -0.00300 &     None     &     None     &     None     &     None     &     None     &     None     &     None     &     None     &     None     &     None     \\
             &              &     None     &     None     &    131.78623 &     None     &    133.23041 &    133.94417 &     None     &    133.88182 &    134.26363 &    134.46326 \\
     0.00000 &      0.00000 &     None     &     None     &     None     &     None     &     None     &     None     &     None     &     None     &     None     &     None     \\
             &              &     None     &     None     &    125.79630 &     None     &    129.94127 &    132.04770 &     None     &    131.82295 &    132.95215 &    133.52334 \\
     0.00000 &      0.00300 &     None     &     None     &     None     &     None     &     None     &     None     &     None     &     None     &     None     &     None     \\
             &              &     None     &     None     &    118.10358 &     None     &    125.52115 &    129.31889 &     None     &    128.98354 &    131.04809 &    132.13809 \\
     0.00300 &     -0.00300 &     None     &     None     &     None     &     None     &     None     &     None     &     None     &     None     &     None     &     None     \\
             &              &     None     &     None     &    127.37315 &     None     &    130.89006 &    132.84476 &     None     &    132.47080 &    133.50836 &    133.94133 \\
     0.00300 &      0.00000 &     None     &     None     &     None     &     None     &     None     &     None     &     None     &     None     &     None     &     None     \\
             &              &     None     &     None     &    113.82811 &     None     &    122.99727 &    129.00451 &     None     &    127.50826 &    130.83289 &    132.06453 \\
     0.00300 &      0.00300 &     None     &     None     &     None     &     None     &     None     &     None     &     None     &     None     &     None     &     None     \\
             &              &     None     &     None     &    102.72718 &     None     &    115.57032 &    123.97104 &     None     &    122.35322 &    127.21930 &    129.41673\\
\hline
\end{tabular}
\caption{Calculated energy spectrum in the pure vector limit of the mixed potential energies. Energy eigenvalues are denoted with two quantum numbers $n$ and $l$ as $E_{nl}$. The units of  $\delta$ and $\lambda b$ are $MeV^{-1}$. The energy eigenvalue is in  $MeV$ unit.}
\label{PVtable}
\end{sidewaystable}

\newpage
\begin{sidewaystable}[tbp]
\begin{tabular}{rrrrrrrrrrrr}
\hline
  $\delta$   & $\lambda b$  &   $E_{00}$   &   $E_{10}$   &   $E_{11}$   &   $E_{20}$   &   $E_{21}$   &   $E_{22}$   &   $E_{30}$   &   $E_{31}$   &   $E_{32}$   &   $E_{33}$   \\
\hline
    -0.00300 &     -0.00300 &   -112.14923 &   -127.07890 &   -130.31170 &   -130.97926 &   -132.26062 &   -133.14380 &   -132.56605 &   -133.20127 &   -133.69216 &   -134.01646 \\
             &              &    127.33455 &    132.98452 &    134.02566 &    134.07068 &    134.44112 &    134.63118 &    134.46021 &    134.63349 &    134.73678 &    134.80009 \\
    -0.00300 &      0.00000 &    -97.84172 &   -119.53740 &   -124.15598 &   -126.39841 &   -128.40409 &   -130.24408 &   -129.50566 &   -130.55972 &   -131.61308 &   -132.39564 \\
             &              &    117.23009 &    129.98704 &    132.36231 &    132.63318 &    133.49989 &    134.01007 &    133.61571 &    134.02737 &    134.30500 &    134.47994 \\
    -0.00300 &      0.00300 &    -87.73377 &   -112.80110 &   -117.75088 &   -121.77591 &   -124.07708 &   -126.64236 &   -126.17297 &   -127.44184 &   -128.95466 &   -130.21501 \\
             &              &    107.10831 &    126.36678 &    130.01599 &    130.78578 &    132.16035 &    133.09651 &    132.49330 &    133.15825 &    133.66895 &    134.00335 \\
     0.00000 &     -0.00300 &   -119.14454 &   -130.18354 &   -132.42563 &   -132.68033 &   -133.52057 &   -134.01920 &   -133.63229 &   -134.03603 &   -134.30943 &   -134.48239 \\
             &              &    119.14454 &    130.18354 &    132.42563 &    132.68033 &    133.52057 &    134.01920 &    133.63229 &    134.03603 &    134.30943 &    134.48239 \\
     0.00000 &      0.00000 &   -105.71706 &   -124.55269 &   -128.49478 &   -129.60941 &   -131.19239 &   -132.38590 &   -131.70348 &   -132.49571 &   -133.15986 &   -133.61163 \\
             &              &    105.71706 &    124.55269 &    128.49478 &    129.60941 &    131.19239 &    132.38590 &    131.70348 &    132.49571 &    133.15986 &    133.61163 \\
     0.00000 &      0.00300 &    -95.22484 &   -118.90859 &   -123.76368 &   -126.17664 &   -128.25226 &   -130.15812 &   -129.40852 &   -130.48890 &   -131.56907 &   -132.36880 \\
             &              &     95.22484 &    118.90859 &    123.76368 &    126.17664 &    128.25226 &    130.15812 &    129.40852 &    130.48890 &    131.56907 &    132.36880 \\
     0.00300 &     -0.00300 &   -127.33455 &   -132.98452 &   -134.02566 &   -134.07068 &   -134.44112 &   -134.63118 &   -134.46021 &   -134.63349 &   -134.73678 &   -134.80009 \\
             &              &    112.14923 &    127.07890 &    130.31170 &    130.97926 &    132.26062 &    133.14380 &    132.56605 &    133.20127 &    133.69216 &    134.01646 \\
     0.00300 &      0.00000 &   -117.23009 &   -129.98704 &   -132.36231 &   -132.63318 &   -133.49989 &   -134.01007 &   -133.61571 &   -134.02737 &   -134.30500 &   -134.47994 \\
             &              &     97.84172 &    119.53740 &    124.15598 &    126.39841 &    128.40409 &    130.24408 &    129.50566 &    130.55972 &    131.61308 &    132.39564 \\
     0.00300 &      0.00300 &   -107.10831 &   -126.36678 &   -130.01599 &   -130.78578 &   -132.16035 &   -133.09651 &   -132.49330 &   -133.15825 &   -133.66895 &   -134.00335 \\
             &              &     87.73377 &    112.80110 &    117.75088 &    121.77591 &    124.07708 &    126.64236 &    126.17297 &    127.44184 &    128.95466 &    130.21501\\
\hline
\end{tabular}
\caption{Calculated energy spectrum in the pure scalar limit of the mixed potential energies. Energy eigenvalues are denoted with two quantum numbers $n$ and $l$ as $E_{nl}$. The units of  $\delta$ and $\lambda b$ are $MeV^{-1}$. The energy eigenvalue is in  $MeV$ unit.}
\label{PStable}
\end{sidewaystable}

\end{document}